\documentclass[10pt,english,fleqn]{article}
\usepackage{newcent}

\usepackage[T1]{fontenc}
\usepackage[utf8]{inputenc}
\usepackage[a4paper]{geometry}
\usepackage{amsmath}
\usepackage{setspace}
\usepackage{amssymb}
\usepackage{esint}

\makeatletter

\usepackage{babel}

\makeatother

\begin{document}
\begin{center}
\textbf{\Large Hamilton-Jacobi formalism for Linearized Gravity}{\Large \vspace{1cm}
}
\par\end{center}{\Large \par}

\begin{center}
M. C. Bertin%
\footnote{mario.bertin@ufabc.edu.br%
}, B. M. Pimentel%
\footnote{pimentel@ift.unesp.br%
}, C. E. Valcárcel%
\footnote{valcarcel@ift.unesp.br%
}, G. E. R. Zambrano%
\footnote{gramos@udenar.edu.co%
}\vspace{0.5cm}

\par\end{center}

\begin{center}
\emph{$^{1}$CMCC, Universidade Federal do ABC.}\\
 \emph{Rua Santa Adália, 166, Santo André, SP, Brazil.} 
\par\end{center}

\begin{center}
\emph{$^{2,3}$Instituto de Física Teórica, UNESP - São Paulo State
University,}\\
 \emph{P. O. Box}\textbf{\emph{ }}\emph{70532-2, 01156-970, São
Paulo, SP, Brazil.} 
\par\end{center}

\begin{center}
\emph{$^{4}$Departamento de Física, Universidad de Nariño,}\\
 \emph{Calle 18 Carrera 50, San Juan de Pasto, Nariño, Colombia.}\vspace{0.25cm}
 \thispagestyle{empty} 
\par\end{center}
\begin{abstract}
In this work we study the theory of linearized gravity via the Hamilton-Jacobi
formalism. We make a brief review of this theory and its Lagrangian
description, as well as a review of the Hamilton-Jacobi approach for
singular systems. Then we apply this formalism to analyze the constraint
structure of the linearized gravity in instant and front-form dynamics.

\vspace{0.5cm}

\end{abstract}

\section{Introduction\label{sec:I}}

$ $

Einstein's field equations in vacuum arise from a variational principle,
setting to zero the first variation of the Einstein-Hilbert action\begin{equation}
S=\frac{1}{2\kappa}\int d^{4}x\sqrt{-g}R,\label{eq:01}\end{equation}
with respect to the metric of space-time, where $R$ is the Ricci's
scalar. The constant $\kappa=8\pi Gc^{-4}$ is obtained in four dimensions
in the weak field approximation. Despite that General Relativity (GR)
has a major difference to other fields, since it treats the gravitational
phenomena as manifestations of the geometry of the space-time, it
has been handled with the same tools for it's canonical quantization.
However, GR as well as the other fundamental interactions is a constrained
theory which requires consistent methods of constraint analysis.

In 1950 Dirac was outlining his Hamiltonian formalism for singular
systems \cite{dirac1}. Studying the gravitational field \cite{dirac5},
he found that a foliation of the space-time simplifies the constraint
structure of gravity with the cost of abandoning the four-symmetry
of the Lagrangian stage.

From a particle physicist's point of view, it would be extremely useful
to have a theory of gravity in a flat space-time that maintains all
the characteristics of the gravitational phenomena in a non-relativistic
limit. This imposition leads us to consider massless fields with spin
0 or 2 (higher even spin fields will only be considered if the spin
2 fails describing the theory). A model of scalar gravitational field
was proposed by Nordström \cite{nordstrom}, but it ended to be in
contradiction with experimentation, since it does not interact with
photons. It also failed when trying to compute the Mercury's perihelion.

The simplest description of gravity as a spin 2 field is the one with
a massless symmetric tensor of rank 2. This model is well described
by the Fierz-Pauli Lagrangian density \cite{fierz}, which becomes
more successful when experimental confrontation comes about. Another
spin 2 field in a fixed background is obtained by linearization of
the GR in the weak-metric approximation, resulting in the linearized
GR (LGR). In this scheme, the linearized Einstein's equations possess
a gauge invariance, and we can use this symmetry to build a Lagrangian
density that describes LGR as a gauge theory. Surprisingly, we obtain
a one-parameter family of Lagrangian densities where the Fierz-Pauli
Lagrangian appears as one of them \cite{padmanabhan}.

Moreover, linearized gravity appears as an attempt to achieve a perturbative
canonical quantization of gravity \cite{dirac5}. At principle, since
these models are based on gauge invariant actions, they are good theories
for the quantization programme proposed by Dirac. However, these theories
still present some difficult problems, e.g. non-renormalizability
in four dimensions (see \cite{kiefer} and references therein). On
the other hand, attempts to learn key properties about quantum gravity
are taken in modified models in two and three dimensions, where the
theories become not only renormalizable, but at least in the two dimensional
case exactly solvable \cite{2D}. In three dimensions, GR is usually
modified with a topological Chern-Simons term \cite{3DTMG}, and more
recently with a massive higher derivative term \cite{3DNMG}. In these
cases, the linearized theories are equivalent to massive Fierz-Pauli
theories, and can be used, for example, to calculate one-loop partition
functions \cite{Dima1}.

On the other hand, there is an increasing interest in field theories
in front-form dynamics \cite{9a}. This kind of dynamics reduces the
number of independent degrees of freedom, which is due to the fact
that the stability group of the Poincaré group in front-form has seven
generators, one more than in the instant-form description. Besides,
the algebra of these these generators takes its simplest form in front-form
dynamics. For some important systems this feature is responsible for
a complete separation of physical degrees of freedom, resulting in
an excitation-free quantum vacuum. This is actually verified, e.g.,
in QCD \cite{QCD} and spontaneous symmetry breaking models \cite{10}.

In this work we study the constraint structure of linearized gravity
in instant and front-form dynamics. For this task, we employ the Hamilton-Jacobi
(HJ) approach for singular systems, first developed by Güler \cite{guler1},
as a generalization of Carathéodory's method for regular mechanics
\cite{caratheodory}. Unlike Dirac's approach \cite{dirac1}, which
is a consistency method to build a Hamiltonian dynamics from a Lagrangian
system, the HJ theory is a full formalism by itself. As necessary
conditions for the existence of extremes of a given action, e.g. \eqref{eq:01},
the constraints of a theory appear as first-order partial differential
equations, whose characteristics equations describe a system with
several independent variables, or parameters. To be sufficient conditions
as well, the so called HJ partial differential equations (PDE) must
also obey integrability, i.e., they must form a complete set of involutive
constraints.

The search for integrability, which is in fact the constraint analysis
by itself, generally reveals two types of HJ equations, called involutive
and non-involutive constraints. Involutive HJ equations are the ones
that form a closed set of integrable equations. The presence of a
non-involutive set indicates dependence between the parameters of
the theory: they must be treated with a redefinition of the phase-space
dynamics. In this context, it is shown in \cite{pimentel4}, for first-order
actions, that the structure of generalized brackets (GB) appears naturally.
Later, a more complete analysis of non-involutive constraints shows
that the GB is a general structure \cite{pimentel6}. Several developments
and applications on the HJ formalism can be found in \cite{pimentel2,baleanu,muslih}.

Our main goal in studying the instant and front-form dynamics of the
LGR is to obtain the algebra of the involutive constraints. In instant-form
there are only involutive constraints, but in front-form the structure
of the dynamics in the coordinates of the light cone reveals a set
of non-involutive HJ equations. This structure allows us to use the
method developed in \cite{pimentel6} to obtain the generalized brackets,
which is an essential tool for canonical quantization.

The paper is structured as follows. Section \ref{sec:HJF} contains
a brief review of the HJ formalism. In section \ref{sec:LG} we introduce
the linearization of the sourceless Einstein's field equations and
its relation to the Fierz-Pauli Lagrangian. Then, we employ the HJ
formalism to integrability analysis, first in the instant-form dynamics
(section \ref{sec:IF}), next in the front-form dynamics (section
\ref{sec:FF}). The last section is dedicated to final remarks.

\section{The Hamilton-Jacobi formalism\label{sec:HJF}}

$ $

Let us consider a Lagrangian function $L(x^{i},\dot{x}^{i},t)$, $i=1,2,...,N$,
whose Hessian matrix\begin{equation}
W_{ij}=\frac{\partial^{2}L}{\partial\dot{x}^{i}\partial\dot{x}^{j}}\label{f01}\end{equation}
is singular of rank $P$. This means that we have $P$ conjugated
momenta\begin{equation}
p_{a}=\frac{\partial L}{\partial\dot{x}^{a}},\ \ a=1,\ldots,P\label{f02}\end{equation}
that can be inverted in relations of the type $\dot{x}^{a}=\dot{x}^{a}\left(p,x,t\right)$,
but $R=N-P$ relations between the canonical variables\begin{equation}
p_{z}+H_{z}=0,\ \ z=1,\ldots,R,\label{f03}\end{equation}
where $H_{z}=-\partial L/\partial\dot{x}^{z}$, correspond to canonical
constraints.

The HJ equation derived from the stationary action principle with
help of Carathéodory's \cite{caratheodory} equivalent Lagrangian
method has the form\begin{equation}
p_{0}+p_{a}\dot{x}^{a}+p_{z}\dot{x}^{z}-L=0,\label{f04}\end{equation}
where $p_{0}\equiv\partial_{t}S$, $p_{a}=\partial_{a}S$, and $p_{z}=\partial_{z}S$.
We may define the canonical Hamiltonian as\begin{equation}
H_{0}\equiv p_{a}\dot{x}^{a}+p_{z}\dot{x}^{z}-L,\label{f05}\end{equation}
then we have a set of $R+1$ Hamilton-Jacobi partial differential
equations (HJ PDE)\begin{equation}
H'_{\alpha}\equiv p_{\alpha}+H_{\alpha}=0,\ \ \alpha=0,1,\ldots,R,\label{f06}\end{equation}
here $x^{0}=t$, and the $H'_{\alpha}$ are just called the Hamiltonian
functions of the theory. In other words, the HJ approach replaces
the study of $R$ canonical constraints with the analysis of $R+1$
HJ PDE.

Being a first-order system, we may use Cauchy's method to solve the
HJ PDE, which gives us a set of total differential equations (TDE)
related to them. The resultant equations are called characteristics
equations,\begin{subequations}\label{008} \begin{gather}
dx^{i}=\frac{\partial H'_{0}}{\partial p_{i}}dx^{0}+\frac{\partial H'_{z}}{\partial p_{i}}dx^{z}=\frac{\partial H'_{\alpha}}{\partial p_{i}}dt^{\alpha},\label{f07}\\
dp_{i}=-\frac{\partial H'_{0}}{\partial x^{i}}dx^{0}-\frac{\partial H'_{z}}{\partial x^{i}}dx^{z}=-\frac{\partial H'_{\alpha}}{\partial x^{i}}dt^{\alpha},\label{f08}\\
dS=p_{a}dx^{a}-H_{\alpha}dt^{\alpha},\label{f09}\end{gather}
\end{subequations}where we have written $t^{\alpha}\equiv(x^{0},x^{z})$
as the independent variables, or parameters, while we see that $(x^{a},p^{a})$
are the dependent variables of the theory.

For any function $F=F(t^{\alpha},x^{a},p^{a})$ we have that\begin{equation}
dF=\frac{\partial F}{\partial x^{a}}dx^{a}+\frac{\partial F}{\partial p^{a}}dp^{a}+\frac{\partial F}{\partial t^{\alpha}}dt^{\alpha}=\{F,H'_{\alpha}\}dt^{\alpha},\label{f10}\end{equation}
where we have used \eqref{f07} and \eqref{f08}, as well as the extended
Poison Brackets\begin{equation}
\{F,G\}\equiv\frac{\partial F}{\partial x^{i}}\frac{\partial G}{\partial p_{i}}-\frac{\partial G}{\partial x^{i}}\frac{\partial F}{\partial p_{i}}+\frac{\partial F}{\partial t}\frac{\partial G}{\partial p_{0}}-\frac{\partial G}{\partial t}\frac{\partial F}{\partial p_{0}}\ .\label{f11}\end{equation}

Let us define a vector field $X_{\alpha}$ such that for any function
$F$ defined in the phase space, $X_{\alpha}(F)\equiv\{F,H'_{\alpha}\}$.
The characteristic equations for the canonical variables can be written
as\begin{equation}
dz^{K}=\{z^{K},H'_{\alpha}\}dt^{\alpha}=X_{\alpha}(z^{K})dt^{\alpha},\label{f12}\end{equation}
where $z^{K}=(x^{i},p^{i})$.

The conditions that ensures the integrability of the system are the
Frobenius' integrability conditions (IC), which are given by $\{H'_{\alpha},H'_{\beta}\}=0$.
On the other hand, these IC imply $[X_{\alpha},X_{\beta}]=0$, i.e.,
the vector fields $X_{\alpha}$ must form a complete orthogonal basis
on the vector space of the parameter space. Generally, Hamiltonians
that obey the Lie algebra $\{H'_{\alpha},H'_{\beta}\}=C_{\alpha\beta}^{\gamma}H'_{\gamma}$
are sufficient to assure integrability \cite{Mishchenko}. However,
these IC imply\begin{equation}
[X_{\alpha},X_{\beta}]F=C_{\beta\alpha}^{\gamma}X_{\gamma}(F)+\{F,C_{\beta\alpha}^{\gamma}\}H'_{\gamma}.\label{f15}\end{equation}
 If the structure coefficients $C_{\alpha\beta}^{\gamma}$ are field
independent, the Lie algebra of the Hamiltonians is reflected in a
Lie algebra of the vector fields. This is sufficient to assure the
existence of a finite Lie group of transformations generated by $X_{\alpha}$.
Otherwise, if the $C_{\alpha\beta}^{\gamma}$ are field dependent,
the last term on the right hand side of \eqref{f15} spoils the algebra
of the vector fields, therefore, the existence of a finite group of
transformations cannot be ensured.

The analysis of IC can also be achieved through the fundamental differential
\eqref{f10}, since\begin{equation}
dH'_{\alpha}=\{H'_{\alpha},H'_{\beta}\}dt^{\beta}=0.\label{f16}\end{equation}
If a subset of Hamiltonians does not satisfy \eqref{f16}, they are
non-involutive constraints, and we may apply the procedure outlined
in \cite{pimentel6}, defining the matrix $M$ with elements $M_{xy}=\{H'_{x},H'_{y}\}$.
If this matrix has rank $S\leq R$, we define the GB with the largest
regular sub-matrix $M_{\bar{a}\bar{b}}=\{H'_{\bar{a}},H'_{\bar{b}}\}$.
In this case, there is an inverse $(M^{-1})^{\bar{a}\bar{b}}$ which
is used to define the Generalized Brackets (GB)\begin{equation}
\{F,G\}^{*}\equiv\{F,G\}-\{F,H'_{\bar{a}}\}(M^{-1})^{\bar{a}\bar{b}}\{H'_{\bar{b}},G\}.\label{f17}\end{equation}
This expression has all the properties of the PB: it is a bilinear
antisymmetric operator that obeys the Jacobi identity and the Leibniz
rule. With the GB the dynamics is given by\begin{equation}
dF=\{F,H'_{\bar{\alpha}}\}^{*}dt^{\bar{\alpha}},\ \ \bar{\alpha}=0,S+1,\ldots,R.\label{f18}\end{equation}
The dynamical evolution of the system depends on $(R-S)$ parameters.
If the system is not complete, new HJ PDE may be found by $\{H'_{\bar{z}},H'_{0}\}=0$,
where $\bar{z}=S+1,\cdots,R$, and IC must be tested for these new
constraints as well.

\section{The Linearized Gravity\label{sec:LG}}

$ $

The linearized General Relativity is obtained from the weak field
approximation of the Einstein's equations\begin{equation}
R^{\mu\nu}-\frac{1}{2}g^{\mu\nu}R=\frac{8\pi G}{c^{4}}\Theta_{\mu\nu},\label{eq:01-1}\end{equation}
where $\Theta_{\mu\nu}$ is the source energy momentum tensor. Here
we decompose the metric $g_{\mu\nu}$ into a Minkowski background
$\eta_{\mu\nu}$, and a perturbation $\phi_{\mu\nu}$,\begin{equation}
g_{\mu\nu}=\eta_{\mu\nu}+\varepsilon\phi_{\mu\nu}+O(\varepsilon^{2}),\label{eq:02}\end{equation}
where $\varepsilon$ is a small parameter introduced to maintain the
correct order of the expansion series. For the LGR only linear terms
in $\varepsilon$ are considered. Under this assumptions and considering
a sourceless gravitational field we obtain, from \eqref{eq:02} in
\eqref{eq:01-1},\begin{equation}
\eta^{\alpha\beta}\partial_{\gamma}\partial^{\gamma}\phi_{\ \nu}^{\nu}-\partial_{\gamma}\partial^{\gamma}\phi^{\alpha\beta}+\partial^{\alpha}\partial_{\lambda}\phi^{\lambda\beta}+\partial^{\beta}\partial_{\lambda}\phi^{\lambda\alpha}-\partial^{\alpha}\partial^{\beta}\phi_{\ \lambda}^{\lambda}-\eta^{\alpha\beta}\partial_{\gamma}\partial_{\mu}\phi^{\mu\gamma}=0.\label{eq03}\end{equation}

On the other hand, \eqref{eq03} can be obtained as the Euler-Lagrange
(EL) equations for the Fierz-Pauli Lagrangian density \cite{fierz}\begin{equation}
\mathcal{L}=\frac{1}{4}\partial_{\mu}\phi_{\ \nu}^{\nu}\partial^{\mu}\phi_{\ \lambda}^{\lambda}-\frac{1}{4}\partial_{\lambda}\phi_{\mu\nu}\partial^{\lambda}\phi^{\mu\nu}+\frac{1}{2}\partial_{\mu}\phi_{\ \nu}^{\mu}\partial_{\lambda}\phi^{\lambda\nu}-\frac{1}{2}\partial_{\mu}\phi^{\mu\nu}\partial_{\nu}\phi_{\lambda}^{\,\,\lambda}.\label{eq:04}\end{equation}
It can be verified that \eqref{eq:04} is invariant under the gauge
transformation\begin{equation}
\phi_{\alpha\beta}\rightarrow\phi_{\alpha\beta}+\partial_{\alpha}\Lambda_{\beta}+\partial_{\beta}\Lambda_{\alpha},\label{eq:05}\end{equation}
where $\Lambda_{\alpha}=\Lambda_{\alpha}(x)$ are arbitrary differentiable
functions. The transformation \eqref{eq:05} is actually similar to
the given in the electromagnetic field. In order to eliminate the
ambiguity raised for this gauge symmetry it is customary to define
a traceless tensor\begin{equation}
h_{\mu\nu}\equiv\phi_{\mu\nu}-\frac{1}{2}\eta_{\mu\nu}\phi_{\ \alpha}^{\alpha},\label{eq:06}\end{equation}
which simplifies \eqref{eq03}:\begin{equation}
\partial^{\mu}\partial_{\mu}h_{\alpha\beta}-\partial^{\mu}\partial_{\alpha}h_{\beta\mu}-\partial^{\mu}\partial_{\beta}h_{\alpha\mu}+\eta_{\alpha\beta}\partial^{\mu}\partial^{\nu}h_{\mu\nu}=0.\label{eq:07}\end{equation}

More important, \eqref{eq:06} allows us to choose\begin{equation}
\partial^{\mu}\partial_{\mu}\Lambda_{\alpha}=-\partial^{\mu}h_{\alpha\mu},\label{eq:08}\end{equation}
from where we obtain a gauge condition\begin{equation}
\partial^{\mu}h_{\alpha\mu}=0,\label{eq:09}\end{equation}
in analogy with the Lorenz gauge from electrodynamics. Equation \eqref{eq:09}
is called de Donder gauge, or harmonic gauge. Finally, the equation
of motion for $h_{\alpha\beta}$ is\begin{equation}
\partial^{\mu}\partial_{\mu}h_{\alpha\beta}=0,\label{eq:10}\end{equation}
which is a relativistic wave equation for a massless spin 2 field,
the graviton. In the linear approximation, the graviton is the mediator
of the gravitational interaction, analogous to the photon which is
the mediator in QED theory. The analysis of the plane wave solution
of \eqref{eq:10}, the polarization states and helicity of the graviton
can be found in \cite{kiefer}.

On the other hand, \eqref{eq:04} is not the only Lagrangian density
for the LGR. There is a one-parameter family of Lagrangians \cite{padmanabhan}
that results in the same field equations \eqref{eq03}. In the next
sections we work only with the Fierz-Pauli Lagrangian \eqref{eq:04}.
In the context of Dirac's formalism in front-form dynamics, this model
was studied in \cite{evens}.

\section{LGR in instant-form\label{sec:IF}}

$ $

The procedure adopted in the preceding section is valid in four dimensions,
but it can be easily extended for $d$ dimensions. We adopt the mostly
minus metric $\eta_{\mu\nu}=diag(+---...)$. Breaking the covariance
in the Lagrangian formalism, making explicit the time variable $\tau=x^{0}$,
we get the Lagrangian density\begin{eqnarray}
\mathcal{L} & = & -\frac{1}{2}\partial_{i}\phi_{i0}\partial_{0}\phi_{00}+\left[\frac{1}{2}\partial_{i}\phi_{00}+\partial_{j}\phi_{ij}-\frac{1}{2}\partial_{i}\phi_{jj}\right]\partial_{0}\phi_{0i}\nonumber \\
 &  & +\left[\frac{1}{4}\delta_{ij}\partial_{0}\phi_{kk}-\frac{1}{4}\partial_{0}\phi_{ij}-\frac{1}{2}\delta_{ij}\partial_{k}\phi_{0k}\right]\partial_{0}\phi_{ij}-\mathcal{V},\label{hj01}\end{eqnarray}
where\begin{eqnarray}
\mathcal{V} & = & \frac{1}{2}\phi_{00}\left[\partial_{i}\partial_{i}\phi_{jj}-\partial_{i}\partial_{j}\phi_{ij}\right]+\frac{1}{2}\phi_{0i}\left[\partial_{i}\partial_{j}\phi_{0j}-\partial_{j}\partial_{j}\phi_{0i}\right]\nonumber \\
 &  & -\frac{1}{4}(\partial_{i}\phi_{jk})^{2}+\frac{1}{4}(\partial_{i}\phi_{jj})^{2}+\frac{1}{2}(\partial_{i}\phi_{ij})^{2}-\frac{1}{2}\partial_{i}\phi_{ij}\partial_{j}\phi_{kk}.\label{hj02}\end{eqnarray}

Due to the symmetry of the field $\phi_{\mu\nu}$ we have that\begin{equation}
\frac{\partial\phi_{\mu\nu}\left(x\right)}{\partial\phi_{\alpha\beta}\left(y\right)}\equiv\Delta_{\alpha\beta}^{\mu\nu}\delta^{d}\left(x-y\right)=\frac{1}{2}\left[\delta_{\alpha}^{\mu}\delta_{\beta}^{\nu}+\delta_{\beta}^{\mu}\delta_{\alpha}^{\nu}\right]\delta^{d}\left(x-y\right),\label{hj03}\end{equation}
where $\delta^{d}\left(x-y\right)$ is Dirac's delta function in $d$
dimensions. The conjugated momenta are given by\begin{subequations}\label{0035}\begin{gather}
p^{00}=-\frac{1}{2}\partial_{i}\phi_{i0},\label{hj04}\\
p^{0i}=\frac{1}{4}\partial_{i}\phi_{00}+\frac{1}{2}\partial_{j}\phi_{ij}-\frac{1}{4}\partial_{i}\phi_{jj},\label{hj05}\\
p^{ij}=\frac{1}{2}\delta_{ij}\partial_{0}\phi_{kk}-\delta_{ij}\partial_{k}\phi_{0k}-\frac{1}{2}\partial_{0}\phi_{ij}.\label{hj06}\end{gather}
\end{subequations}This system is singular, and we identify equations
\eqref{hj04} and \eqref{hj05} as constraints.

It was pointed out by Anderson \cite{anderson} that it is possible
to simplify the canonical constraints. Particularly, we may simplify
calculations by adding surface terms in the Lagrangian, with the identity\begin{equation}
\partial_{\rho}\phi_{\alpha\beta}\partial_{\gamma}\phi_{\mu\nu}=\partial_{\gamma}\phi_{\alpha\beta}\partial_{\rho}\phi_{\mu\nu}+\partial_{\rho}(\phi_{\alpha\beta}\partial_{\gamma}\phi_{\mu\nu})-\partial_{\gamma}(\phi_{\alpha\beta}\partial_{\rho}\phi_{\mu\nu}).\label{hj07}\end{equation}
Then we are able to eliminate the dependence in $\partial_{0}\phi_{0\mu}$
and obtain\begin{equation}
\mathcal{L}=\left(\partial_{i}\phi_{0j}-\delta_{ij}\partial_{k}\phi_{0k}+\frac{1}{4}\delta_{ij}\partial_{0}\phi_{kk}-\frac{1}{4}\partial_{0}\phi_{ij}\right)\partial_{0}\phi_{ij}-\mathcal{V}.\label{hj08}\end{equation}
The new conjugated momenta are\begin{subequations}\label{0036}\begin{gather}
\pi^{0\mu}=0,\label{hj09}\\
\pi^{ij}=\frac{1}{2}\delta_{ij}\partial_{0}\phi_{kk}-\frac{1}{2}\partial_{0}\phi_{ij}+\frac{1}{2}\partial_{i}\phi_{0j}+\frac{1}{2}\partial_{j}\phi_{0i}-\delta_{ij}\partial_{k}\phi_{0k}.\label{hj10}\end{gather}
\end{subequations}We have reduced the constraints \eqref{hj04} and
\eqref{hj05} in one single constraint \eqref{hj09}. This fact has
a close resemblance with the electromagnetic case, where the primary
constraint has the form $\pi^{0}=0$. Equation \eqref{hj10} is a
dynamical relation, from where we get the velocities as functions
of the conjugated momenta\begin{equation}
\partial_{0}\phi_{ij}=-2\pi^{ij}+\frac{2}{(d-2)}\delta_{ij}\pi^{kk}+\partial_{i}\phi_{0j}+\partial_{j}\phi_{0i}.\label{hj11}\end{equation}
We notice that \eqref{hj10} is not defined in two dimensions. The
canonical Hamiltonian density is given by\begin{eqnarray}
\mathcal{H}_{0} & = & -\left(\pi^{ij}\right)^{2}+\frac{1}{(d-2)}\left(\pi^{kk}\right)^{2}-2\phi_{0i}\mathcal{C}^{i}+\frac{1}{2}\phi_{00}\mathcal{C}^{0}\nonumber \\
 &  & -\frac{1}{4}\left(\partial_{i}\phi_{jk}\right)^{2}+\frac{1}{4}\left(\partial_{i}\phi_{jj}\right)^{2}+\frac{1}{2}\left(\partial_{i}\phi_{ij}\right)^{2}-\frac{1}{2}\partial_{i}\phi_{ij}\partial_{j}\phi_{kk},\label{hj12}\end{eqnarray}
where we define the functions\begin{subequations}\label{HJ13}\begin{gather}
\mathcal{C}^{0}\equiv\partial_{i}\partial_{i}\phi_{jj}-\partial_{i}\partial_{j}\phi_{ij},\label{hj13}\\
\mathcal{C}^{i}\equiv\partial_{j}\pi^{ij}.\label{hj14}\end{gather}
\end{subequations}

In the context of the HJ formalism, we have $(d+1)$ Hamiltonian densities\begin{subequations}\label{HJ15}\begin{gather}
\mathcal{H}'^{\tau}=\pi^{\tau}+\mathcal{H}_{0}=0,\label{hj15}\\
\mathcal{H}'^{0\mu}=\pi^{0\mu}=0.\label{hj16}\end{gather}
\end{subequations}The first relation is related to the time variable
$\tau=x^{0}$, while the second one is related to the variables $\phi_{0\mu}$,
that now stands as parameters of the theory. The fundamental PB, observing
\eqref{hj03}, are given by\begin{equation}
\{\phi_{\alpha\beta}(x),\pi^{\mu\nu}(y)\}=\Delta_{\alpha\beta}^{\mu\nu}\ \delta^{d-1}\left(\mathbf{x}-\mathbf{y}\right).\label{hj18}\end{equation}
All PB are computed at equal times $x^{0}=y^{0}=cte.$

The characteristics equations of the theory suggest the definition
of the fundamental differential\begin{equation}
dF=\{F,\mathcal{H}'^{\tau}\}d\tau+\{F,\mathcal{H}'^{00}\}d\phi_{00}+2\{F,\mathcal{H}'^{0i}\}d\phi_{0i},\label{hj19}\end{equation}
where integration is implicit on the right hand side. The factor $2$
in the last term is due to the symmetry of $\phi_{\mu\nu}$.

Following the next step in the HJ formalism, we test the integrability
conditions for the Hamiltonian densities. We obtain\begin{subequations}\label{HJ20}\begin{gather}
d\mathcal{H}'^{00}=-\frac{1}{2}\,\mathcal{C}^{0}d\tau=0,\label{hj20}\\
d\mathcal{H}'^{0i}=\mathcal{C}^{i}d\tau=0.\label{hj21}\end{gather}
\end{subequations}Then, $\mathcal{C}^{\mu}$ defined in \eqref{HJ13}
are new Hamiltonian densities, corresponding to the HJ equations $\mathcal{C}^{\mu}=0$,
and the IC have to be tested with them as well. From these new densities,
the only non-zero PB is\begin{equation}
\{\mathcal{C}'^{0}(x),\mathcal{H}'^{\tau}(y)\}=\partial_{i}\partial_{j}\pi^{ij}\ \delta^{d-1}\left(\mathbf{x}-\mathbf{y}\right)=\partial_{i}\mathcal{C}^{i}\ \delta^{d-1}\left(\mathbf{x}-\mathbf{y}\right).\label{hj22}\end{equation}
This means that the IC for these Hamiltonian densities are identically
satisfied and the system is considered complete.

Once we have the complete set of Hamiltonian densities \eqref{HJ13}
and \eqref{HJ15}, we are able to build the evolution of the system
with the differential\begin{equation}
dF=\{F,\mathcal{H}'^{\tau}\}d\tau+\{F,\mathcal{H}'^{00}\}d\phi_{00}+2\{F,\mathcal{H}'^{0i}\}d\phi_{0i}+\{F,\mathcal{C}^{\mu}\}d\omega_{\mu},\label{hj23}\end{equation}
where $\omega_{\mu}$ are new parameters related to the Hamiltonians
$\mathcal{C}^{\mu}$. Again, integration is implicit on the right
side. The complete set of Hamiltonian densities is in involution,
i.e, the PB are identically zero or they are linear combinations of
the previous Hamiltonian densities. In particular, the algebra of
the generators $\mathcal{H}'^{0\mu}$ and $\mathcal{C}^{\mu}$ is
abelian.

For this involutive system, the characteristic equations are given
by \eqref{hj23}. For $F=\phi_{\mu\nu}$ we have\begin{eqnarray}
d\phi_{\mu\nu} & = & \left[-2\Delta_{\mu\nu}^{ij}\pi^{ij}+\frac{2}{d-2}\Delta_{\mu\nu}^{ii}\pi^{jj}+2\Delta_{\mu\nu}^{ij}\partial_{i}\phi_{0j}\right]d\tau\nonumber \\
 &  & +\Delta_{\mu\nu}^{00}d\phi_{00}+2\Delta_{\mu\nu}^{0i}d\phi_{0i}-\Delta_{\mu\nu}^{ij}\partial_{i}d\omega_{j}.\label{hj25}\end{eqnarray}
These equations reproduce the fact that $\phi_{0\mu}$ are parameters
of the theory, since their velocities cannot be fixed $\left(d\phi_{0\mu}=d\phi_{0\mu}\right)$.
They also give us back the relation \eqref{hj11}, as expected, apart
of the term in $\omega_{j}$.

For $F=\pi^{\mu\nu}$, we obtain\begin{eqnarray}
d\pi^{\mu\nu} & = & \left[\frac{1}{2}\Delta_{jj}^{\mu\nu}\left(\partial_{i}\partial_{i}\phi_{kk}-\partial_{i}\partial_{i}\phi_{00}-\partial_{i}\partial_{k}\phi_{ik}\right)\right.\nonumber \\
 &  & +\frac{1}{2}\Delta_{ij}^{\mu\nu}\left(\partial_{i}\partial_{j}\phi_{00}-\partial_{i}\partial_{j}\phi_{kk}-\partial_{k}\partial_{k}\phi_{ij}+2\partial_{i}\partial_{k}\phi_{kj}\right)\nonumber \\
 &  & \left.+\frac{1}{2}\Delta_{00}^{\mu\nu}\left(\partial_{i}\partial_{j}\phi_{ij}-\partial_{i}\partial_{i}\phi_{jj}\right)+2\Delta_{0i}^{\mu\nu}\partial_{j}\pi^{ij}\right]d\tau\nonumber \\
 &  & +\left[\Delta_{ij}^{\mu\nu}\partial_{i}\partial_{j}-\Delta_{jj}^{\mu\nu}\partial_{i}\partial_{i}\right]d\omega_{0}.\label{hj26}\end{eqnarray}
They reproduce the EL equations \eqref{eq03} apart of the linear
term in $\omega_{0}$, as follows: the equation for $\pi^{00}$ is
equivalent to the first IC \eqref{hj13}, which is also the EL equation
\eqref{eq03} with $\alpha=\beta=0$. For $\pi^{0i}$ the correspondent
characteristic equation is equivalent to the second IC \eqref{hj14},
and gives the EL equation for $\alpha=0$ and $\beta=i$. The dynamical
equations of the theory are actually the equations for $\pi^{ij}$,
which became the EL equation for $\alpha=i$ and $\beta=j$, when
\eqref{hj10} is taken account. Then, the characteristics equations
are equivalent to the EL equations when appropriate parameters $\omega_{\mu}$
are chosen.

\section{LGR in front-form\label{sec:FF}}

$ $

In relativistic field theories we are free to choose the parameter
that determines the time evolution. This freedom comes from the physical
requirement of Poincaré covariance. When dealing with a field theory
in flat space-time, the choice of a particular parameter $\tau$ comes
with the choice of a family of surfaces $\Sigma_{\tau}=constant$.
If we knew the configuration of the fields over one of the members
of the family the field equations in canonical form should give us
the evolution of this configuration on later surfaces in a unique
way. It was outlined by Dirac \cite{nullplane} that the quantization
of a relativistic field theory in instant-form is not the only kind
of relativistic dynamics. In fact there are at least five inequivalent
forms of Hamiltonian dynamics of relativistic field theories \cite{bekker}.
One of them is the front-form dynamics.

If we have a $d$-dimensional Minkowski space-time, the light-cone
coordinates are defined by\begin{subequations}\label{LG01}\begin{gather}
x^{+}=\frac{1}{\sqrt{2}}(x^{0}+x^{d-1}),\label{lg01}\\
x^{-}=\frac{1}{\sqrt{2}}(x^{0}-x^{d-1}),\label{lg02}\\
x^{i}=x^{i}\ ,\ i=1,2,...,d-2.\label{lg03}\end{gather}
\end{subequations}In this, we set $\tau=x^{+}$ as the new time parameter,
and $x^{-}$ and $x^{i}$ stands as spatial coordinates. The transverse
coordinates are denoted by $\mathbf{x}=(x^{1},...,x^{n})$, with $n=d-2$.
Therefore, the dynamics of fields in this coordinate system is given
by the configuration over a surface $x^{+}=\tau_{0}$ and its evolution
to later surfaces by means of a Hamiltonian function. This kind of
dynamics is often called front-form, null-plane, or even light-front
dynamics, and the surfaces of constant $x^{+}$ are called null-planes.
Since a null-plane divides space-like and time-like vectors, the causal
structure is included into the light-cone coordinates.

In order to obtain the conjugated momenta, we will separate the time
and spatial coordinates from the Lagrangian density:\begin{eqnarray}
\mathcal{L} & = & \partial_{+}\phi_{++}\left[-\frac{1}{2}\partial_{-}\phi_{--}\right]+\partial_{+}\phi_{+-}\left[-\frac{1}{2}\partial_{-}\phi_{ii}\right]+\partial_{+}\phi_{+i}\partial_{-}\phi_{i-}\nonumber \\
 &  & +\partial_{+}\phi_{--}\left[\frac{1}{2}\partial_{-}\phi_{++}-\partial_{i}\phi_{i+}+\frac{1}{2}\partial_{+}\phi_{ii}\right]\nonumber \\
 &  & +\partial_{+}\phi_{-i}\left[-\frac{1}{2}\partial_{+}\phi_{-i}+\partial_{k}\phi_{ki}+\partial_{i}\phi_{+-}-\frac{1}{2}\partial_{i}\phi_{kk}\right]\nonumber \\
 &  & +\partial_{+}\phi_{ij}\left[-\frac{1}{2}\delta_{ij}\partial_{-}\phi_{+-}+\frac{1}{2}\delta_{ij}\partial_{-}\phi_{kk}-\frac{1}{2}\delta_{ik}\partial_{-}\phi_{kj}-\frac{1}{2}\delta_{ij}\partial_{k}\phi_{k-}\right]-\mathcal{V},\label{lg04}\end{eqnarray}
where\begin{eqnarray}
\mathcal{V} & = & \phi_{++}\left[\frac{1}{2}\partial_{i}\partial_{i}\phi_{--}+\frac{1}{2}\partial_{-}\partial_{-}\phi_{ii}-\partial_{-}\partial_{i}\phi_{i-}\right]\nonumber \\
 &  & +\phi_{+-}\left[-\frac{1}{2}\partial_{i}\partial_{i}\phi_{+-}+\partial_{i}\partial_{i}\phi_{kk}+\partial_{-}\partial_{i}\phi_{+i}-\partial_{i}\partial_{k}\phi_{ik}\right]\nonumber \\
 &  & +\phi_{+i}\left[-\partial_{k}\partial_{k}\phi_{-i}-\frac{1}{2}\partial_{-}\partial_{-}\phi_{+i}+\partial_{-}\partial_{k}\phi_{ki}+\partial_{i}\partial_{k}\phi_{k-}-\partial_{-}\partial_{i}\phi_{kk}\right]\nonumber \\
 &  & +\left[\frac{1}{2}\partial_{i}\phi_{im}\partial_{k}\phi_{km}-\frac{1}{2}\partial_{i}\phi_{ik}\partial_{k}\phi_{mm}+\frac{1}{4}(\partial_{i}\phi_{kk})^{2}-\frac{1}{4}(\partial_{i}\phi_{km})^{2}\right].\label{lg05}\end{eqnarray}

As we did in instant-form, we may perform partial integrations and
eliminate surface terms in order to simplify the expressions for the
momenta, obtaining the equivalent Lagrangian density\begin{eqnarray}
\mathcal{L} & = & \partial_{+}\phi_{--}\left[-\partial_{i}\phi_{i+}+\frac{1}{2}\partial_{+}\phi_{ii}\right]\nonumber \\
 &  & +\partial_{+}\phi_{-i}\left[-\frac{1}{2}\partial_{+}\phi_{-i}+\partial_{k}\phi_{ki}+\partial_{i}\phi_{+-}+\partial_{-}\phi_{+i}\right]\nonumber \\
 &  & +\partial_{+}\phi_{ij}\left[-\delta_{ij}\partial_{-}\phi_{+-}+\frac{1}{2}\delta_{ij}\partial_{-}\phi_{kk}-\frac{1}{2}\partial_{-}\phi_{ij}-\delta_{ij}\partial_{k}\phi_{-k}\right]-\mathcal{V}.\label{lg06}\end{eqnarray}
From here we may write the momenta\begin{subequations}\label{LG07}\begin{gather}
\pi^{+\mu}=0\ ,\label{lg07}\\
\pi^{--}=\frac{1}{2}\partial_{+}\phi_{ii}-\partial_{i}\phi_{+i}\ ,\label{lg08}\\
\pi^{-i}=\frac{1}{2}\left(\partial_{-}\phi_{+i}-\partial_{+}\phi_{-i}+\partial_{k}\phi_{ik}+\partial_{i}\phi_{+-}\right)\ ,\label{lg09}\\
\pi^{ij}=\frac{1}{2}\delta_{ij}\partial_{+}\phi_{--}-\delta_{ij}\partial_{-}\phi_{+-}+\frac{1}{2}\delta_{ij}\partial_{-}\phi_{kk}-\frac{1}{2}\partial_{-}\phi_{ij}-\delta_{ij}\partial_{k}\phi_{-k}.\label{lg10}\end{gather}
\end{subequations}

Relations \eqref{lg08} and \eqref{lg09} can be inverted to obtain
the velocities\begin{subequations}\label{LG11}\begin{gather}
\partial_{+}\phi_{ii}=2\pi^{--}+2\partial_{i}\phi_{+i}\label{lg11}\\
\partial_{+}\phi_{-i}=-2\pi^{-i}+\partial_{-}\phi_{+i}+\partial_{k}\phi_{ik}+\partial_{i}\phi_{+-}.\label{lg12}\end{gather}
\end{subequations}Relation \eqref{lg10} has a peculiarity. The trace
part can be inverted to obtain\begin{equation}
\partial_{+}\phi_{--}=\frac{2}{n}\pi^{ii}-\left(\frac{n-1}{n}\right)\partial_{-}\phi_{ii}+2(\partial_{-}\phi_{+-}+\partial_{i}\phi_{-i}),\label{lg13}\end{equation}
for $n\neq0$ ($d\neq2$). The traceless part, on the other hand,
is a constraint\begin{equation}
\bar{\pi}^{ij}+\frac{1}{2}\partial_{-}\bar{\phi}_{ij}=0.\label{lg14}\end{equation}
Here, the bar on any tensor is defined by\begin{equation}
\bar{A}_{ij}\equiv A_{ij}-\frac{1}{n}\delta_{ij}A_{kk},\label{lg15}\end{equation}
which describes its traceless part. We notice that for the four dimensional
case, i.e. $n=2$, $\bar{\phi}_{ij}=h_{ij}$. Now we compute the canonical
Hamiltonian density:\begin{eqnarray}
\mathcal{H}_{\tau} & = & 2\pi^{-i}\left[\partial_{j}\bar{\phi}_{ij}+\frac{1}{n}\partial_{i}\phi_{kk}-\pi^{-i}\right]+\pi^{--}\left[\frac{2}{n}\pi^{kk}-\frac{n-1}{n}\partial_{-}\phi_{kk}+2\partial_{k}\phi_{-k}\right]\nonumber \\
 &  & -\phi_{++}\mathcal{C}^{+}-2\phi_{+-}\mathcal{C}^{-}-2\phi_{+i}\mathcal{C}^{i}-\frac{1}{4}(\partial_{i}\bar{\phi}_{jk})^{2}\nonumber \\
 &  & -\frac{1}{2}\partial_{i}\bar{\phi}_{ij}\partial_{j}\phi_{kk}+\frac{n-3}{4n}(\partial_{i}\phi_{jj})^{2},\label{lg16}\end{eqnarray}
where\begin{subequations}\label{LG17}\begin{gather}
\mathcal{C}^{+}=\partial_{i}\partial_{-}\phi_{i-}-\frac{1}{2}\partial_{-}\partial_{-}\phi_{ii}-\frac{1}{2}\partial_{i}\partial_{i}\phi_{--},\label{lg17}\\
\mathcal{C}^{-}=\partial_{i}\pi^{-i}+\partial_{-}\pi^{--}-\frac{1}{2}\partial_{i}\partial_{i}\phi_{jj},\label{lg18}\\
\mathcal{C}^{i}=\partial_{-}\pi^{-i}+\frac{1}{n}\partial_{i}\left[\pi^{kk}-\frac{1}{2}\partial_{-}\phi_{kk}\right]-\partial_{-}\partial_{j}\bar{\phi}_{ij}+\frac{1}{2}\partial_{i}\partial_{k}\phi_{-k}+\frac{1}{2}\partial_{k}\partial_{k}\phi_{-i}.\label{lg19}\end{gather}
\end{subequations}

Following the HJ formalism we have the Hamiltonian densities\begin{subequations}\label{LG20}\begin{gather}
\mathcal{H}'^{\tau}\equiv\pi^{\tau}+\mathcal{H}_{\tau}=0,\label{lg20}\\
\mathcal{H}'^{+\mu}\equiv\pi^{+\mu}=0,\label{lg21}\\
\mathcal{Q}'^{ij}\equiv\bar{\pi}^{ij}+\frac{1}{2}\partial_{-}\bar{\phi}_{ij}=0.\label{lg22}\end{gather}
\end{subequations}The first equation is related to the time parameter,
the second to the $\phi_{+\mu}$ fields, and the last one to the traceless
part of $\phi_{ij}$. From these densities we identify the parameters
of the theory and build the fundamental differential\begin{eqnarray}
dF & = & \{F,\mathcal{H}'^{\tau}\}d\tau+\{F,\mathcal{H}'^{++}\}d\phi_{++}\nonumber \\
 &  & +2\{F,\mathcal{H}'^{+-}\}d\phi_{+-}+2\{F,\mathcal{H}'^{+i}\}d\phi_{+i}+\{F,\mathcal{Q}'^{ij}\}d\bar{\phi}_{ij}.\label{lg23}\end{eqnarray}
As usual, integration is implicit on the right hand side.

Now we proceed testing integrability and searching for new Hamiltonian
densities. We obtain\begin{subequations}\label{LG24}\begin{gather}
d\mathcal{H}'^{++}=\mathcal{C}^{+}d\tau=0,\label{lg24}\\
d\mathcal{H}'^{+-}=\mathcal{C}^{-}d\tau=0,\label{lg25}\\
d\mathcal{H}'^{+i}=\mathcal{C}^{i}d\tau=0,\label{lg25a}\end{gather}
\end{subequations}that identifies $\mathcal{C}^{+}$, $\mathcal{C}^{-}$,
and $\mathcal{C}^{i}$ as new Hamiltonian densities of the system.
We may write\begin{equation}
\mathcal{C}^{i}=\partial_{-}\pi^{-i}+\partial_{j}\pi^{ij}+\frac{1}{2}\partial_{i}\partial_{k}\phi_{-k}+\frac{1}{2}\partial_{k}\partial_{k}\phi_{-i},\label{lg26}\end{equation}
where we have made a simplification with help of Hamiltonian \eqref{lg22}.
The IC $d\mathcal{Q}'^{ij}=0$ will give a relation between the parameters
$\tau=x^{+}$ and $\bar{\phi}_{ij}$. This means that these parameters
are not independent, and we must eliminate this dependence with apropriate
GB. Testing the integrability of the generators $\mathcal{C}^{\mu}$
we may see that there are no more Hamiltonians, then the system is
considered completed.

For each density \eqref{LG20}, we have related an independent variable
$\left(\tau,\phi_{+\mu},\bar{\phi}_{ij}\right)$. However, for the
densities \eqref{LG17} we have to add a new set of variables, $\left(\omega_{+},\omega_{-},\omega_{i}\right)$
respectively, to the theory. Therefore, we define the new fundamental
differential\begin{eqnarray}
dF & = & \{F,\mathcal{H}'^{\tau}\}d\tau+\{F,\mathcal{H}'^{++}\}d\phi_{++}+2\{F,\mathcal{H}'^{+-}\}d\phi_{+-}\nonumber \\
 &  & +2\{F,\mathcal{H}'^{+i}\}d\phi_{+i}+\{F,\mathcal{Q}'^{ij}\}d\bar{\phi}_{ij}+\{F,\mathcal{C}^{\mu}\}d\omega_{\mu}.\label{lg28}\end{eqnarray}

With the purpose of reducing the phase space with only the independent
parameters of the theory, we have to analyze the algebra of the Hamiltonian
densities. We have that $\mathcal{H}'^{+\mu}$ and $\mathcal{C}^{\mu}$
are in involution. On the other hand, the non-involutive Hamiltonian
density $\mathcal{Q}'^{ij}$ satisfies\begin{equation}
\{\mathcal{Q}^{ij}(x),\mathcal{Q}^{ij}(y)\}=P^{ijkl}\partial_{-}\delta(x^{-}-y^{-})\delta^{n}(\mathbf{x}-\mathbf{y}),\label{lg29}\end{equation}
where $P^{ijkl}$ is a projector tensor, since it projects any transverse
tensor of rank 2 in its symmetric traceless part \begin{equation}
P^{ijkl}\equiv\Delta_{kl}^{ij}-\frac{1}{n}\delta_{ij}\delta_{kl}.\label{lg30}\end{equation}
This projector is not defined in the two dimensional case.

As we have mentioned, the parameters related to the non-involutive
constraints can be eliminated of the dynamical evolution after we
compute the GB. We start by building the matrix\begin{eqnarray}
M^{\left(ij,kl\right)}(x,y) & \equiv & \{\mathcal{Q}^{ij}(x),\mathcal{Q}^{kl}(y)\}\ .\label{lg31}\end{eqnarray}
The inverse is given by\begin{equation}
(M^{-1})_{\left(ij,kl\right)}(x,y)=\frac{1}{2}W_{ijkl}\epsilon(x^{-}-y^{-})\delta^{n}(\mathbf{x}-\mathbf{y})+f_{ijkl},\label{lg33}\end{equation}
where $\epsilon(x)$ is the step function and $W_{ijkl}$ is the inverse
of the projector $P^{ijkl}$:\begin{equation}
W_{ijkl}=\frac{n}{n-1}P^{ijkl},\label{lg35}\end{equation}
which satisfies $P^{ijmn}W_{mnkl}=\Delta_{kl}^{ij}$. The existence
of $W_{ijkl}$ is assured for $n>1$, since we can verify that $P^{ijkl}$
is a regular matrix in this case.

The $f_{ijkl}$ are arbitrary functions that do not depend on $x^{-}$.
They appear as consequence of the null-plane dynamics because we have
not specified sufficient boundary conditions to uniquely determine
the evolution of the system \cite{steinhardt}. Therefore, this inverse
is not unique, but represents a family of matrices. It is possible
to determine boundary conditions such that the boundary terms are
zero, and a unique dynamics emerges. This behavior is characteristic
of the front-form dynamics, as outlined in \cite{Cas}. Let us make
$f_{ijkl}=0$, in this case the GB can be defined as\begin{gather}
\{F(x),G(y)\}^{*}\equiv\{F(x),G(y)\}\nonumber \\
\,\,\,\,\,\,\,\,\,\,\,\,\,\,\,\,\,\,\,\,\,\,\,\,\,\,\,\,\,\,\,\,\,\,-\int dz\int dw\{F(x),\mathcal{Q}^{ij}(z)\}(M^{-1})_{\left(ij,kl\right)}(z,w)\{\mathcal{Q}^{kl}(w),G(y)\},\label{lg36}\end{gather}
and the fundamental GB are\begin{subequations}\label{LG37}\begin{gather}
\left\{ \phi_{\mu\nu},\phi_{\alpha\beta}\right\} ^{*}=\frac{1}{2}\Delta_{\mu\nu}^{ij}\Delta_{\alpha\beta}^{kl}P^{ijkl}\epsilon\left(x^{-}-y^{-}\right)\delta^{n}\left(\mathbf{x}-\mathbf{y}\right),\label{lg37}\\
\left\{ \phi_{\mu\nu},\pi^{\alpha\beta}\right\} ^{*}=\left[\Delta_{\mu\nu}^{\alpha\beta}+\frac{1}{2}\Delta_{\mu\nu}^{ij}\Delta_{kl}^{\alpha\beta}P^{ijkl}\right]\delta\left(x^{-}-y^{-}\right)\delta^{n}\left(\mathbf{x}-\mathbf{y}\right),\label{lg38}\\
\left\{ \pi^{\mu\nu},\pi^{\alpha\beta}\right\} ^{*}=\frac{1}{2}\Delta_{ij}^{\mu\nu}\Delta_{kl}^{\alpha\beta}P^{ijkl}\partial_{-}\delta\left(x^{-}-y^{-}\right)\delta^{n}\left(\mathbf{x}-\mathbf{y}\right).\label{lg39}\end{gather}
\end{subequations}By direct calculation, we see that these GB applied
to the constraints of the theory result in a closed algebra and, therefore,
all constraints become involutive: integrability is then achieved.
Besides, the algebra of the involutive constraints $ $$\mathcal{H}'^{+\mu}$
and $\mathcal{C}^{\mu}$ is abelian indeed. This is expected since
we need the algebra and the number of involutive constraints of a
relativistic theory to be independent of the choice of dynamics for
a good dynamical description. This ensures that all time-preserved
quantities are also independent of this choice.

With the GB, the dynamics of the system is given by the differential\begin{equation}
dF=\{F,\mathcal{H}'^{\tau}\}^{*}d\tau+\{F,\mathcal{H}'^{+\mu}\}^{*}d\phi_{+\mu}+\{F,\mathcal{C}^{\mu}\}^{*}d\omega_{\mu}.\label{lg40}\end{equation}
Then we may express the characteristics equations of the system. Let
us begin with the variables $\phi_{\mu\nu}$:\begin{subequations}\label{LG41}\begin{gather}
d\phi_{+\mu}=d\phi_{+\mu},\label{lg41}\\
d\phi_{--}=\left[\frac{2}{n}\pi^{kk}-\left(\frac{n-1}{n}\right)\partial_{-}\phi_{kk}+2\partial_{k}\phi_{-k}+2\partial_{-}\phi_{+-}\right]d\tau-\partial_{-}d\omega_{-},\label{lg42}\\
d\phi_{-i}=\left[-2\pi^{-i}+\partial_{j}\phi_{ij}+\partial_{i}\phi_{+-}+\partial_{-}\phi_{+i}\right]d\tau-\frac{1}{2}\left(\partial_{i}d\omega_{-}+\partial_{-}d\omega_{i}\right).\label{lg43}\end{gather}
\end{subequations}The first equation is expected, since the variables
$\phi_{+\mu}$ are parameters related to the Hamiltonians $\mathcal{H}'^{+\mu}$.
Equations \eqref{lg42} and \eqref{lg43} are equivalent to \eqref{lg13}
and \eqref{lg12} with proper choice of the parameters $\omega_{-}$
and $\omega_{i}$. For the equation of the trace of $\phi_{ij}$ we
obtain\begin{equation}
d\phi_{ii}=\left[2\pi^{--}+2\partial_{m}\phi_{+m}\right]d\tau-\partial_{i}d\omega_{i},\label{lg44}\end{equation}
which is just equal to equation (\ref{lg11}) if we set $\partial_{i}d\omega_{i}=0$.
For $i\neq j$ we have\begin{eqnarray}
d\bar{\phi}_{ij} & = & \frac{1}{2}P^{ijkl}d\tau\int dy^{-}d^{n}y\epsilon\left(x^{-}-y^{-}\right)\delta^{n}\left(x-y\right)\times\nonumber \\
 &  & \times\left[2\partial_{-}\partial_{k}\phi_{+l}-2\partial_{k}\pi^{-l}+\frac{1}{2}\partial_{k}\partial_{l}\phi_{mm}+\frac{1}{2}\partial_{m}\partial_{m}\phi_{kl}\right]\nonumber \\
 &  & -\frac{1}{2}\left(\partial_{i}d\omega_{j}+\partial_{j}d\omega_{i}\right).\label{lg45}\end{eqnarray}
This is actually a dynamical equation. With some work it is possible
to show that this is the equivalent EL equation \eqref{eq03} for
$\left(\alpha,\beta\right)=\left(i,j\right)$, with $i\neq j$.

For the momenta, we have the relations\begin{subequations}\label{LG46}\begin{gather}
d\pi^{++}=\left[\partial_{i}\partial_{-}\phi_{i-}-\frac{1}{2}\partial_{i}\partial_{i}\phi_{--}-\frac{1}{2}\partial_{-}\partial_{-}\phi_{ii}\right]d\tau,\label{lg46}\\
d\pi^{+-}=\left[\partial_{i}\pi^{-i}+\partial_{-}\pi^{--}-\frac{1}{2}\partial_{i}\partial_{i}\phi_{kk}\right]d\tau,\label{lg47}\\
d\pi^{+i}=\left[\partial_{-}\pi^{-i}+\frac{1}{n}\partial_{i}\pi^{kk}-\partial_{-}\partial_{j}\phi_{ij}+\frac{1}{2n}\partial_{i}\partial_{-}\phi_{kk}\right.\nonumber \\
\,\,\,\,\,\,\,\,\,\,\,\,\,\,\,\,\,\,\,\left.+\frac{1}{2}\left(\partial_{i}\partial_{k}\phi_{-k}+\partial_{k}\partial_{k}\phi_{-i}\right)\right]d\tau.\label{lg48}\end{gather}
\end{subequations}These equations represent the integrability conditions
that give rise to the constraints $\mathcal{C}'^{\mu}=0$. They are
the non-dynamical set of EL equations.

The following equations\begin{subequations}\label{LG49}\begin{gather}
d\pi^{--}=-\frac{1}{2}\partial_{k}\partial_{k}\phi_{++}d\tau+\frac{1}{2}\partial_{k}\partial_{k}d\omega_{+}\label{lg49}\\
d\pi^{-i}=\left[\partial_{i}\pi^{--}-\frac{1}{2}\partial_{i}\partial_{-}\phi_{++}+\frac{1}{2}\left(\partial_{i}\partial_{j}+\delta_{j}^{i}\partial_{k}\partial_{k}\right)\phi_{+j}\right]d\tau\nonumber \\
\,\,\,\,\,\,\,\,\,\,\,\,\,\,\,\,\,\,\,-\frac{1}{2}\partial_{i}\partial_{-}d\omega_{+}-\frac{1}{4}\left(\partial_{i}\partial_{j}+\delta_{j}^{i}\partial_{k}\partial_{k}\right)d\omega_{j}\label{lg50}\\
d\pi^{ij}=\left[\frac{1}{2}\partial_{j}\pi^{-i}+\frac{1}{2}\partial_{i}\pi^{-j}+\frac{1}{n}\delta_{ij}\partial_{k}\pi^{-k}-\frac{1}{2}\partial_{-}\partial_{j}\phi_{+i}-\frac{1}{2}\partial_{-}\partial_{i}\phi_{+j}\right.\nonumber \\
\,\,\,\,\,\,\,\,\,\,\,\,\,\,\,\,\,\,\,\left.-\frac{1}{n}\delta_{ij}\partial_{-}\partial_{k}\phi_{+k}-\frac{1}{4}\partial_{k}\partial_{k}\phi_{ij}-\frac{1}{4}\partial_{i}\partial_{j}\phi_{kk}+\frac{1}{2n}\delta_{ij}\partial_{k}\partial_{k}\phi_{ll}\right]d\tau\nonumber \\
\,\,\,\,\,\,\,\,\,\,\,\,\,\,\,\,\,\,\,-\frac{1}{8}\partial_{-}\partial_{j}d\omega_{i}-\frac{1}{8}\partial_{-}\partial_{i}d\omega_{j}+\frac{1}{4n}\delta_{ij}\partial_{-}\partial_{k}d\omega_{k},\label{lg51}\end{gather}
\end{subequations}complete the remaining set of EL equations.

\section{Final Remarks\label{sec:Final-Remarks}}

$ $

In this work we have used the HJ formalism to analyze the constraints
of linearized gravity. We found that, while the instant-form dynamics
have only constraints in involution, a sub-set of Hamiltonian densities
in the front-form dynamics are non-involutive. The later case becomes
a good laboratory to build the GB in the context of the HJ formalism.

We have carried out the usual procedure of construction of a Lagrangian
density from the properties of gauge invariance of the linearized
Einstein's equations. In both forms of dynamics, we were able to modify
the Lagrangian in order to obtain simplifications on the momenta,
and therefore to analyze the structure of their Hamiltonian densities.
Using the IC, we were able to find the complete set of Hamiltonian
densities.

In instant-form, the theory has constraints that come from the IC,
represented by the Hamiltonian densities \eqref{HJ13}. Together with
\eqref{HJ15}, they form a complete integrable set. In particular,
the densities \eqref{HJ13} close an abelian Lie algebra with the
Poisson brackets. To build the field equations, we have extended the
space of parameters to embrace the independent variables related to
the Hamiltonians \eqref{HJ13}. The analysis resulted to be in full
accordance with the field equations \eqref{eq03}.

In the front-form dynamics, we have found a richer structure. There
was a subset of non-involutive constraints, represented by the densities
\eqref{lg22}. With this set we have built the GB, eliminating the
traceless variables $\bar{\phi}_{ij}$. As usual when describing a
theory in the coordinates of the light-cone, these GB are unique only
if boundary conditions are carefully chosen on a null-plane $x^{-}=cte$,
setting to zero the arbitrary functions $f_{ijkl}$ that appear in
\eqref{lg33}.

We also verified that the involutive constraints $ $$\mathcal{H}'^{+\mu}$
and $\mathcal{C}^{\mu}$ obey an abelian Lie algebra, this time with
respect to the generalized brackets \eqref{lg36}. This is despite
the fact that the Hamiltonians $\mathcal{C}^{i}$ do not close an
algebra with the Poisson brackets. It is a very good feature of the
front-form description of this theory that the non-involutive constraints
$\mathcal{Q}^{ij}$ are exactly those needed to ensure the correct
algebra of these constraints via the definition \eqref{lg36}. It
can be seen that, for the computation of $\left\{ \mathcal{C}^{i},\mathcal{C}^{j}\right\} ^{*}$,
the second term in the right side of \eqref{lg36} exactly cancels
the non-zero term $ $$\left\{ \mathcal{C}^{i},\mathcal{C}^{j}\right\} $.
As expected, because of the presence of the constraints $\mathcal{C}^{\mu}=0$,
the characteristics equations of the system have arbitrary parameters
$\omega^{\mu}$ not related to variables of the system. The fundamental
differential \eqref{lg40} is built with the complete set of Hamiltonian
functions, and gives rise to characteristics equations that are again
equivalent to the field equations \eqref{eq03}.

\section*{Acknowledgments}

\begin{spacing}{1.3}
MCB thanks UFABC and FAPESP for partial support. BMP thanks CNPq and
CAPES for partial support. CEV thanks CAPES for full support. GERZ
thanks VIPRI-UDENAR for full support. 
\end{spacing}

\end{document}